\documentclass[conference]{IEEEtran}

\usepackage{cite}
\usepackage{amsmath,amssymb,amsfonts}
\usepackage{multirow}
\usepackage{graphicx}
\usepackage{textcomp}
\usepackage{xcolor}
\usepackage{url}
\usepackage{verbatim}
\usepackage{flushend}

\usepackage{listings}
\usepackage[linesnumbered,boxed]{algorithm2e}
\SetAlCapSkip{1em}
\usepackage{pgfplots}
\usepackage{tikz}
\usetikzlibrary{backgrounds}

\usepackage{color, colortbl}
\definecolor{LightGray}{gray}{0.9}
\definecolor{DarkBlue}{RGB}{33,113,181}
\definecolor{MediumBlue}{RGB}{107,174,214}
\definecolor{LightBlue}{RGB}{158,202,225}
\definecolor{VeryLightBlue}{RGB}{198,219,239}
\definecolor{amaranth}{rgb}{0.9, 0.17, 0.31}
\definecolor{celestialblue}{rgb}{0.29, 0.59, 0.82}
\definecolor{dartmouthgreen}{rgb}{0.05, 0.5, 0.06}

\newtheorem{definition}{Definition}[section]
\newtheorem{theorem}{Theorem}[section]
\newtheorem{example}{Example}[section]

\begin{document}
\title{Order in Desbordante: Techniques for Efficient Implementation of Order Dependency Discovery Algorithms}
\date{}

\author{
\IEEEauthorblockN{Yakov Kuzin, Dmitriy Shcheka, Michael Polyntsov, Kirill Stupakov, Mikhail Firsov, George Chernishev}
\IEEEauthorblockA{Saint-Petersburg University \\ Saint-Petersburg, Russia \\ \{yakov.s.kuzin, dmitriy.v.shcheka, polyntsov.m, kirill.v.stupakov, mikhail.a.firsov, chernishev\}@gmail.com}}
\maketitle

                                                    
\newcommand{\relation}[1]{\textbf{#1}}
\newcommand{\attribute}[1]{\textsf{#1}}
\newcommand{\tuple}[1]{\textsf{\textit{#1}}}
\newcommand{\tupleValue}[2]{\tuple{#1}_\attribute{#2}}

\newcommand{\attributeSet}[1]{\mathcal{#1}}
\newcommand{\emptySet}[0]{\{\}}
\newcommand{\tupleProjectionOnSet}[2]{\tuple{#1}_\attributeSet{#2}}

\newcommand{\attributeList}[1]{\textsf{\textbf{#1}}}
\newcommand{\listPermutation}[1]{\textsf{\textbf{#1}}'}
\newcommand{\emptyList}[0]{[\, ]}
\newcommand{\tupleProjection}[2]{\tuple{#1}_\attributeList{#2}}

\newcommand{\orderDependency}[3]{#1 \mapsto_{#2} #3}
\newcommand{\functionalDependency}[3]{#1 \rightarrow_{#2} #3}

\newcommand{\todo}[1]{\textcolor{red}{[TODO: #1]}}
\newcommand{\blue}[1]{\textcolor{blue}{#1}}

                      
\begin{abstract}
Science-intensive data profiling focuses on discovery and validation of various patterns in datasets. This study considers discovery of one such pattern~--- order dependency (OD). Simply put, OD states that some list of columns is ordered according to another one. It is of use for database query optimization, data cleaning and deduplication, anomaly detection, and much more.

Existing discovery methods have approached this problem solely from the algorithmic standpoint, without focusing on the implementation side. At the same time, this problem is very computationally intensive, and therefore this part should not be ignored, as it brings ODs closer to industrial use.

In this paper, we study two algorithms for OD discovery which target different OD axiomatizations~--- FASTOD and ORDER. We start by reimplementing these algorithms in C++ in order to speed them up and lower their memory consumption. We then analyze their bottlenecks and propose several techniques which improve their performance even further.

To perform evaluation, we have implemented these algorithms inside Desbordante~--- a science-intensive, high-performance, and open-source data profiling tool developed in C++. Experiments have demonstrated a performance improvement of up to 3x obtained by reimplemented versions, and, with the application of our techniques, up to 10x. Memory consumption has been lowered by up to 2.9x.
\end{abstract}

                                                                                               
\section{Introduction}

Currently, there exists a considerable scholarly interest in the analysis of extensive datasets. These datasets often exhibit various inconsistencies, including missing values, duplicates, and many other anomalies~\cite{DBLP:conf/icde/ChuIP13}.

Data profiling~\cite{10.5555/3312004} is a research field that aims to detect and characterize such inconsistencies in order to prepare them for further use (e.g., data cleaning). Data profiling can be divided into two kinds~\cite{DBLP:journals/corr/abs-2301-05965}: naive and science-intensive. The former aims to extract simple dataset characteristics such as number of rows and columns, number of nulls, mean and variance, etc. The science-intensive kind concerns the extraction of complex patterns represented by structures which we will refer to as \textit{primitives}. Examples of such patterns are database dependencies (functional~\cite{DBLP:journals/VLDB/PapenbrockEMNRSZN15}, inclusion~\cite{10.1145/3357384.3357916}), association rules~\cite{10.5555/2677098}, algebraic constraints~\cite{10.5555/1315451.1315509}, inferred semantic data types~\cite{10.1145/3292500.3330993}, and others. These patterns are typically discovered through employing various algorithms, which are  usually very costly, resulting in dataset size being a significant limiting factor. Thus, the development of novel efficient algorithms and improving the performance of existing ones are relevant problems.

One such primitive is the order dependency (OD). Informally, an OD states that some column is ordered according to another column. For example, an increase in salary in the IT department payroll table might be directly correlated with increases in programmer's grade. According to reference~\cite{DBLP:journals/pvldb/SzlichtaGGKS17}, ODs prove effective in improving data quality, as their violation may serve as an indicator of underlying data errors. Furthermore, as discussed in~\cite{DBLP:journals/pvldb/SzlichtaGGZ13}, these dependencies can be leveraged by different database query optimizers to fine-tune query performance.

Desbordante (Spanish for \textit{boundless})~\cite{DBLP:journals/corr/abs-2301-05965} is a \textit{science-intensive}, \textit{high-performance}, and \textit{open-source} data profiling tool implemented in C++. To the best of our knowledge, Desbordante is currently the only profiler that possesses these three qualities. It is capable of discovering and validating many primitives, including functional dependencies (both exact and approximate), conditional functional dependencies, metric functional dependencies, and others. The full list can be found on the web-site (https://github.com/Mstrutov/Desbordante/). However, Desbordante currently lacks support for ODs, which we aim to add.

ODs have been known since 80es~\cite{DBLP:journals/TCS/GinsburgH83}, and therefore this subject contains a vast body of work. In this paper we focus on two recent types of OD, which are based on different axiomatizations~--- list-based~\cite{DBLP:journals/vldb/LangerN16} and set-based~\cite{DBLP:journals/pvldb/SzlichtaGGKS17}. Different formalisms effectively lead to different primitives, each having its own algorithm and resulting primitive instances (result set). The list-based axiomatization offers the ORDER algorithm, while the set-based one~--- FASTOD.

However, existing discovery approaches have considered this problem from the algorithmic standpoint only, without focusing on the implementation side. To evaluate their algorithms, authors have developed research prototypes implemented in Java. Firstly, our previous studies~\cite{9435469} demonstrated that merely reimplementing these algorithms in C++ can improve their run times up to 3.5 times and lower memory consumption up to 2.5 times. Secondly, applying various code-level optimizations~\cite{10143047} can improve run times even further, up to 8 times. As the result, existing implementations are slower than they could be. Since primitive discovery problem is very computationally intensive, the engineering part should not be ignored as it brings ODs closer to industrial use.

In this paper, we develop a technical approach to the problem based on efficient implementations of algorithms for OD discovery. We start by reimplementing these algorithms in C++ in order to speed them up and lower their memory consumption. Then we analyze their bottlenecks and propose several techniques which improve their performance even further.

Overall, the contribution of this paper is the following:
\begin{itemize}
    \item A comprehensive study of two recent formalizations of OD and a description of the algorithms for their discovery~--- ORDER and FASTOD.
    \item Several novel techniques for efficient implementation of both algorithms.
    \item Open-source C++ implementations of both the algorithms and proposed techniques.
    \item Experimental evaluation of the proposed techniques and discussion of the results.
\end{itemize}

This paper is organized as follows. In Section~\ref{sec:background} we provide definitions for both axiomatizations, after which we discuss them with examples. Next, in Section~\ref{sec:relwork} we present related work concerning ODs. In Section~\ref{sec:algos} we describe existing algorithms and our improvements. Their evaluation is discussed in Section~\ref{sec:exper}, and Section~\ref{sec:concl} concludes the paper.

                                                                                     
\section{Background}~\label{sec:background}
Currently, there exist two axiomatizations describing the notion of order dependency. These axiomatizations define different objects, which results in different algorithms for their discovery. The first one treats left hand side and right hand side of a dependency as lists, and the second one~--- as sets of attributes. In order to understand the respective algorithms and our modifications, it is necessary to grasp the basics of these axiomatizations. Therefore, in this section we present essential concepts and formal definitions, closely following~\cite{szlichta2012fundamentals} and \cite{DBLP:journals/pvldb/SzlichtaGGKS17} while presenting descriptive examples.

\subsection{Basic Definitions}

\textbf{Relations.} $\relation{R}$ denotes a \textit{relation (schema)} and $\relation{r}$ denotes a specific \textit{relation instance (table)}. $\attribute{A}$, $\attribute{B}$ and $\attribute{C}$ denote single \textit{attributes}, $\tuple{s}$ and $\tuple{t}$ denote \textit{tuples}, and $\tupleValue{t}{A}$ denotes the value of an attribute $\attribute{A}$ in a tuple $\tuple{t}$.

\textbf{Sets.} $\attributeSet{X}$ and $\attributeSet{Y}$ denote \textit{sets} of attributes. Let $\tupleProjectionOnSet{t}{X}$ denote the \textit{projection} of tuple \tuple{t} on $\attributeSet{X}$. $\attributeSet{X}\attributeSet{Y}$ is a shorthand for $\attributeSet{X} \cup \attributeSet{Y}$. The empty set is denoted as $\emptySet$.

\textbf{Lists.} $\attributeList{X}$, $\attributeList{Y}$ and $\attributeList{Z}$ denote \textit{lists} of attributes. Empty list is denoted as $\emptyList$. $[\attribute{A}, \attribute{B}, \attribute{C}]$ denotes an explicit list. $[\attribute{A} \,\,| \,\, \attributeList{T}]$ denotes a list with \textit{head} $\attribute{A}$ and \textit{tail} $\attributeList{T}$. Let $\attributeList{X}\attributeList{Y}$ be a concatenation of lists $\attributeList{X}$ and $\attributeList{Y}$. Set $\attributeSet{X}$ denotes the set of elements in list $\attributeList{X}$. Any place a set is expected but a list appears, the list is cast to a set; e.g., $\tupleProjection{t}{X}$ denotes $\tupleProjectionOnSet{t}{X}$. Let $\listPermutation{X}$ denote some other permutation of elements of list $\attributeList{X}$.

\vspace{1em}
\begin{definition}
    Given a relational schema $\relation{R}$ and an instance $\relation{r}$ over $\relation{R}$ with attribute sets $\attributeSet{X}, \attributeSet{Y} \subset \relation{R}$, we say that a functional dependency (FD) $\functionalDependency{\attributeSet{X}}{}{\attributeSet{Y}}$ holds iff for any $\tuple{s}, \tuple{t} \in \relation{r}$, the following is true: $\tupleProjectionOnSet{s}{X}=\tupleProjectionOnSet{t}{X} \Rightarrow \tupleProjectionOnSet{s}{Y}=\tupleProjectionOnSet{t}{Y}$.
\end{definition}

\vspace{1em}
\begin{example}
    For instance, in Table~\ref{tab:shipment_table} functional dependency $\functionalDependency{\{\textit{``Shipment cost''}\}}{}{\{\textit{``Weight''}\}}$ holds, since in all tuples with equal values of attribute \textit{``Shipment cost''} ($t_1$ and $t_5$), the values of ``Weight'' are equal as well.
\end{example}

\begin{table}[htp]
    \centering
    \caption{A table with shipment information, adapted from~\cite{DBLP:journals/vldb/LangerN16}}
    \label{tab:shipment_table}    
    \begin{tabular}{|c|c|c|c|c|}
        \hline
        $t_{id}$ & Weight & Distance & Shipment cost & Days \\ \hline \hline
        \rowcolor{LightGray}
        1 & 10 & 40 & 17 & 3 \\ \hline
        2 & 15 & 80 & 48 & 7 \\ \hline
        \rowcolor{LightGray}
        3 & 8 & 60 & 13 & 5 \\ \hline
        4 & 15 & 90 & 28 & 8 \\ \hline
        \rowcolor{LightGray}
        5 & 10 & 40 & 17 & 4 \\ \hline
        6 & 25 & 100 & 43 & 9 \\ \hline
        \rowcolor{LightGray}
        7 & 9 & 60 & 18 & 6 \\ \hline
    \end{tabular}
\end{table}

With FDs, however, it is impossible to capture relationships among ordered attributes, such as timestamps or numbers, which are quite common in business data. Therefore, the concept of OD is introduced, generalizing FD by allowing comparison operators other than $=$.

\vspace{1em}
\begin{definition}
    Let $\attributeList{X}$ be a list of attributes and $\theta \in \{ \le, <, >, \ge \}$. For two tuples $\tuple{r}$ and $\tuple{s}$, $\attributeSet{X} \in \relation{R}$ we say that $\tupleProjection{r}{X} \, \theta \, \tupleProjection{s}{X}$ if
    \begin{enumerate}
        \item $\attributeList{X} = \emptyList$, \textit{or}
        \item $\attributeList{X} = [\attribute{A} \,\,| \,\, \attributeList{T}] \land \tupleValue{r}{A} \, \theta \, \tupleValue{s}{A}$, \textit{or}
        \item $\attributeList{X} = [\attribute{A} \,\,| \,\, \attributeList{T}] \land \tupleValue{r}{A} = \tupleValue{s}{A} \land \tupleProjection{r}{T} \, \theta \, \tupleProjection{s}{T}$.
    \end{enumerate}
    Unless otherwise specified, numbers are ordered numerically, strings are ordered lexicographically and dates are ordered chronologically.
\end{definition}

\vspace{1em}
\begin{definition}
    The order dependency $\orderDependency{\attributeList{X}}{\theta}{\attributeList{Y}}$, where $\theta \in \{ \le, <, >, \ge \}$, is present in the instance $\relation{r}$ over the relation $\relation{R}$ iff for any $\tuple{s}, \tuple{t} \in \relation{r}$, the condition $\tupleProjection{s}{X} \, \theta \, \tupleProjection{t}{X} \Rightarrow \tupleProjection{s}{Y} \, \theta \, \tupleProjection{t}{Y}$ holds. If $\theta$ is omitted, it is implied that $\theta$ is $<$.
\end{definition}

\vspace{1em}
Essentially, the presence of OD $\orderDependency{\attributeList{X}}{}{\attributeList{Y}}$ means that, when ordering the values by $\attributeList{X}$, the resulting list would also be ordered by $\attributeList{Y}$. 

Discovered order dependencies have many applications, such as database query optimization, data cleaning and deduplication, anomaly detection, and much more.

\vspace{1em}
\begin{example}
    In Table~\ref{tab:shipment_table} we can see that the order dependency $\orderDependency{[\textit{``Distance''}, \textit{``Weight''}]}{\leq}{[\textit{``Days''}]}$ holds. Ordering by \textit{``Distance''} and breaking ties by \textit{``Weight''} ($t_1 \leq t_5 \leq t_3 \leq t_7 \leq t_2 \leq t_4 \leq t_6$) is the same as ordering by \textit{``Days''}. Also note that although FD $\functionalDependency{\{\textit{``Weight''}\}}{}{\{\textit{``Shipment cost''}\}}$ holds, as seen in previous example, OD  $\orderDependency{[\textit{``Weight''}]}{\leq}{[\textit{``Shipment cost''}]}$ does not hold. We discuss reasons for this in Section~\ref{sec:background::ssec:violations}. 
\end{example}

\subsection{List-based definitions}

\begin{definition}
    Two attribute lists $\attributeList{X}$ and $\attributeList{Y}$ are \textit{order compatible} with respect to $\theta \in \{\leq, <, >, \geq\}$, denoted as $\attributeList{X} \sim_{\theta} \attributeList{Y}$, iff $\attributeList{X}\attributeList{Y} \leftrightarrow_{\theta} \attributeList{Y}\attributeList{X}$ ($\orderDependency{\attributeList{X}\attributeList{Y}}{\theta}{\attributeList{Y}\attributeList{X}}$ and $\orderDependency{\attributeList{Y}\attributeList{X}}{\theta}{\attributeList{X}\attributeList{Y}}$). $\emptyList$ is order compatible with any attribute list. ODs in the form of $\attributeList{X} \sim_{\theta} \attributeList{Y}$ are called order compatible dependencies (OCDs)
\end{definition}

\vspace{1em}
\begin{example}
    In Table~\ref{tab:shipment_table} OCD $[\textit{``Distance''}] \sim [\textit{"Days"}]$ is valid: sorting by \textit{``Distance''} and breaking ties by \textit{``Days''} is equivalent to sorting by \textit{``Days''} and breaking ties by \textit{``Distance''}.
\end{example}

\vspace{1em}
\begin{definition}
    An attribute list $\attributeList{X}$ is \textit{minimal}, iff for any disjoint, contiguous sub-lists $\attributeList{V}$ and $\attributeList{W}$ in $\attributeList{X}$, such that $\attributeList{W}$ precedes (not necessarily directly) $\attributeList{V}$, OD $\orderDependency{\attributeList{V}}{}{\attributeList{W}}$ does not hold.
\end{definition}

\vspace{1em}
\begin{definition}
    The order dependency $\orderDependency{\attributeList{X}}{}{\attributeList{Y}}$ is \textit{minimal}, iff
    \begin{enumerate}
        \item there is no prefix $\attributeList{V}$ of $\attributeList{X}$, such that $\orderDependency{\attributeList{V}}{}{\attributeList{Y}}$ does not hold, \textit{and}
        \item there is no prefix $\attributeList{W}$ of $\attributeList{Y}$, such that $\orderDependency{\attributeList{X}}{}{\attributeList{W}}$ holds, \textit{and}
        \item $\attributeList{X}$ is minimal, \textit{and}
        \item $\attributeList{Y}$ is minimal.
    \end{enumerate}
\end{definition}

\subsection{Violations} \label{sec:background::ssec:violations}

ODs can be violated in two ways. We begin with the following theorem and then explain how the two conditions therein correspond to two possible sources of violations. Detailed proofs can be found in the original paper~\cite{DBLP:journals/pvldb/SzlichtaGGKS17}.

\vspace{1em}
\begin{theorem}
    For every instance $\relation{r}$ of relation $\relation{R}$ and $\theta \in \{\leq, <, >, \geq\}$, $\orderDependency{\attributeList{X}}{\theta}{\attributeList{Y}} \iff \orderDependency{\attributeList{X}}{\theta}{\attributeList{XY}} \land \attributeList{X} \sim_{\theta} \attributeList{Y}$.
\end{theorem}

\vspace{1em}
\begin{definition}
    Tuples $\tuple{s}$ and $\tuple{t}$ form a \textit{split} with respect to a pair of attribute lists $(\attributeList{X}, \attributeList{Y})$, if $\tupleProjection{s}{X} = \tupleProjection{t}{X}$, but $\tupleProjection{s}{Y} \neq \tupleProjection{t}{Y}$.
\end{definition}

\vspace{1em}
\begin{definition}
    Tuples $\tuple{s}$ and $\tuple{t}$ form a \textit{merge} with respect to a pair of attribute lists $(\attributeList{X}, \attributeList{Y})$, if $\tupleProjection{s}{X} \neq \tupleProjection{t}{X}$, but $\tupleProjection{s}{Y} = \tupleProjection{t}{Y}$.
\end{definition}

\vspace{1em}
A \textit{split} among $(\attributeList{X}, \attributeList{Y})$ implies a \textit{merge} among $(\attributeList{Y}, \attributeList{X})$ and vice versa. Presence of \textit{split} or \textit{merge} implies $\orderDependency{\attributeList{X}}{\theta}{\attributeList{XY}}$ being violated. Introducing both as separate concepts facilitates the distinction between the two types of order dependencies: \textit{splits} invalidate only order dependencies under $\leq$ and $\geq$, and \textit{merges} invalidate only order dependencies under $<$ and $>$.

\vspace{1em}
\begin{definition}
    Tuples $\tuple{s}$ and $\tuple{t}$ form a \textit{swap} with respect to pair of attribute lists $(\attributeList{X}, \attributeList{Y})$ and $\theta \in \{ \leq, <, >, \geq \}$, if $\tupleProjection{s}{X} \, \theta \, \tupleProjection{t}{X}$, but $\neg(\tupleProjection{s}{Y} \, \theta \, \tupleProjection{t}{Y})$. Presence of a \textit{swap} implies $\attributeList{X} \sim_{\theta} \attributeList{Y}$ being violated.
\end{definition}

\vspace{1em}
\begin{example}
    Order dependency in Table~\ref{tab:shipment_table} $\orderDependency{[\textit{``Weight''}]}{\leq}{[\textit{``Shipment cost''}]}$ does not hold because of a split $(t_2, t_4)$. OD $\orderDependency{[\textit{``Days''}]}{}{[\textit{``Shipment cost''}]}$ does not hold because of a swap $(t_1, t_3)$.
\end{example}

\vspace{1em}
\begin{theorem}
    $\orderDependency{\attributeList{X}}{<}{\attributeList{Y}}$ iff $\orderDependency{\attributeList{X}}{\leq}{\attributeList{Y}}$
\end{theorem}

This theorem unifies dependencies under operators $<$ and $\leq$. One of the algorithms discussed in this paper, ORDER~\cite{DBLP:journals/vldb/LangerN16}, uses this fact by discovering only dependencies under strict comparison operators.

\subsection{Set-based definitions}

In~\cite{DBLP:journals/pvldb/SzlichtaGGKS17} a polynomial mapping from list-based representation to a set-based canonical form of ODs is presented, allowing the traversal of a much smaller set-containment lattice instead of a list-containment lattice.

\vspace{1em}
\begin{definition}
    Let $\relation{R}$ be a relation schema, and $\relation{r}$ be its instance. The \textit{equivalence class} of tuple $\tuple{t} \in \relation{r}$ with respect to a given set of attributes $\attributeSet{X}$ is defined as the set $\varepsilon(\tupleProjectionOnSet{t}{X}) = \{\tuple{s} \in \relation{r} \mid \tupleProjectionOnSet{s}{X} = \tupleProjectionOnSet{t}{X}\}$.
\end{definition}

\vspace{1em}
\begin{definition}
    An attribute $\attribute{A}$ is considered \textit{constant} within each equivalence class concerning the set of attributes $\attributeSet{X}$ (denoted as $\attributeSet{X} : \emptyList \rightarrow _{cst} \attribute{A}$) if there exists an order dependency $\orderDependency{\listPermutation{X}}{}{\listPermutation{X}\attribute{A}}$ for any permutation $\listPermutation{X}$ of elements in $\attributeSet{X}$.
\end{definition}

\vspace{1em}
\begin{definition}
    Two attributes $\attribute{A}$ and $\attribute{B}$ are \textit{order compatible} within each equivalence class regarding the set of attributes $\attributeSet{X}$ (denoted as $\attributeSet{X} : \attribute{A} \sim \attribute{B}$) if there exists a permutation $\listPermutation{X}$, such that $\orderDependency{\listPermutation{X}\attribute{A}}{}{\listPermutation{X}\attribute{B}}$.
\end{definition}

\vspace{1em}
\begin{definition}
    Dependencies of the form $\attributeSet{X} : \emptyList \rightarrow _{cst} \attribute{A}$ and  $\attributeSet{X} : \attribute{A} \sim \attribute{B}$ are referred to as \textit{canonical} order dependencies. $\attributeSet{X}$ is called the context.
\end{definition}

\vspace{1em}
In~\cite{DBLP:journals/pvldb/SzlichtaGGKS17} theorems are presented that show a way of mapping list-based OD representations to equivalent set-based canonical forms of ODs. Given a set of attributes $\attributeSet{X}$, for brevity, we state $\forall j, \attribute{Y}_j$ to mean $\forall j \in \{1, 2, \ldots, |\attributeSet{Y}|\}, \attribute{Y}_j$.

\vspace{1em}
\begin{theorem} \label{theorem:mapping_to_set_based}
    $\orderDependency{\attributeList{X}}{}{\attributeList{Y}}$ iff $\forall i, \attributeSet{X} : \emptyList \rightarrow_{cst} \attribute{Y}_i$ and $\forall i, j, \{\attribute{X}_1, \ldots, \attribute{X}_{i-1}, \attribute{Y}_1, \ldots, \attribute{Y}_{j-1}\}: \attribute{X}_i \sim \attribute{Y}_j$
\end{theorem}

\vspace{1em}
\begin{example}
    By theorem~\ref{theorem:mapping_to_set_based}, an OD $\orderDependency{\attribute{A}\attribute{B}}{}{\attribute{C}\attribute{D}}$ can be mapped into the following equivalent set of canonical ODs:
    \begin{enumerate}
        \item $\{\attribute{A}, \attribute{B}\} : \emptyList \rightarrow_{cst} \attribute{C}$,
        \item $\{\attribute{A}, \attribute{B}\} : \emptyList \rightarrow_{cst} \attribute{D}$,
        \item $\emptySet : \attribute{A} \sim \attribute{C}$,
        \item $\{\attribute{A}\} : \attribute{B} \sim \attribute{C}$,
        \item $\{\attribute{C}\} : \attribute{A} \sim \attribute{D}$,
        \item $\{\attribute{A}, \attribute{C}\} : \attribute{B} \sim \attribute{D}$.
    \end{enumerate}
\end{example}

\vspace{1em}
\begin{definition}
    A canonical OD $\attributeSet{X} : \emptyList \rightarrow _{cst} \attribute{A}$ is \textit{trivial}, if $\attribute{A} \in \attributeSet{X}$. A canonical OD $\attributeSet{X} : \attribute{A} \sim \attribute{B}$ is \textit{trivial} if
    \begin{enumerate}
        \item $\attribute{A} \in \attributeSet{X}$, \textit{or}
        \item $\attribute{B} \in \attributeSet{X}$, \textit{or}
        \item $\attribute{A} = \attribute{B}$.
    \end{enumerate}
\end{definition}

\vspace{1em}
\begin{definition}
    A canonical OD  $\attributeSet{X} : \emptyList \rightarrow _{cst} \attribute{A}$ is \textit{minimal} if it is not \textit{trivial} and there is no context $\attributeSet{Y} \subset \attributeSet{X}$, such that $\attributeSet{Y} : \emptyList \rightarrow _{cst} \attribute{A}$ holds. A canonical OD $\attributeSet{X} : \attribute{A} \sim \attribute{B}$ is \textit{minimal} if it is not \textit{trivial} and
    \begin{enumerate}
        \item there is no context $\attributeSet{Y} \subset \attributeSet{X}$, such that $\attributeSet{Y} : \attribute{A} \sim \attribute{B}$ holds, \textit{or}
        \item $\attributeSet{X} : \emptyList \rightarrow _{cst} \attribute{A}$, \textit{or}
        \item $\attributeSet{X} : \emptyList \rightarrow _{cst} \attribute{B}$.
    \end{enumerate}
\end{definition}

Violations for set-based axiomatization are defined similarly.

\subsection{OD discovery algorithms}

In this paper we consider the following OD discovery algorithms:
\begin{enumerate}
    \item ORDER~\cite{DBLP:journals/vldb/LangerN16}, which operates with \textit{list-based} order dependencies,
    \item FASTOD~\cite{DBLP:journals/pvldb/SzlichtaGGKS17}, which works with equivalent \textit{set-based} mapping of \textit{list-based} dependencies.
\end{enumerate}

The set of dependencies produced by FASTOD is proven to be complete~\cite{DBLP:journals/pvldb/SzlichtaGGKS17}. ORDER, however, uses excessively aggressive pruning rules, which leads to the resulting set of dependencies being incomplete in the following ways:
\begin{enumerate}
    \item The algorithm skips dependencies in the form of $\orderDependency{\attributeList{X}}{}{\attributeList{X}\attributeList{Y}}$,
    \item if $\orderDependency{\attributeList{X}}{}{\attributeList{Y}}$ is invalidated by a \textit{swap}, then $\orderDependency{\attributeList{X}\attribute{A}}{}{\attributeList{Y}\attribute{B}}$ is not considered, leading to ODs in the form of $\orderDependency{\attributeList{X}\attribute{A}}{}{\attributeList{X}\attribute{A}\attributeList{Y}\attribute{B}}$ being missed,
    \item if $\orderDependency{\attributeList{X}\attribute{A}}{}{\attributeList{Y}\attribute{B}}$ is invalidated by a \textit{split}, then $\attributeList{X}\attribute{A} \sim \attributeList{Y}\attribute{B}$ is not considered, which maps to set-based canonical OD $\attributeSet{X}\attributeSet{Y}: \attribute{A} \sim \attribute{B}$.
\end{enumerate}

FASTOD also has a more concise way of representing constants, producing only one \textit{set-based} canonical OD $\emptySet : \emptyList \rightarrow _{cst} \attribute{B}$ for each constant, while ORDER produces ODs $\orderDependency{[\attribute{A}]}{}{[\attribute{B}]}$ for all attributes $\attribute{A}$ and all constants $\attribute{B}$.

The fact that ORDER yields incomplete results, however, does not mean that it should not be used for dependency mining. ORDER's aggressive pruning rules allow it to perform better, especially on large datasets, while still being able to give out potentially useful dependencies.

                                                                                                
\section{Related Work}\label{sec:relwork}

According to~\cite{szlichta2012fundamentals}, order dependencies had first been proposed in ~\cite{DBLP:journals/TCS/GinsburgH83}. Since then, a lot of dependency types differing from the initial ones had been discovered. Furthermore, alternative axiomatizations of initial concepts has been proposed: some were characterized by lists of attributes, while others~--- by sets. Pointwise Order Dependencies (PODs) and Lexicographical Order Dependencies (LODs) are examples of dependency classes with different axiomatizations. PODs~--- set-based dependencies~--- were presented by Seymour Ginsburg and Richard Hull~\cite{DBLP:journals/TCS/GinsburgH83} in the context of databases. LODs~--- list-based dependencies and a more useful alternative to PODs due to their applications in query optimization, were studied in~\cite{DBLP:journals/ATDS/Wilfred01}. Jaroslaw Szlichta et al.~\cite{szlichta2012fundamentals} presented the set of list-based inference rules which defines this class. Based on~\cite{szlichta2012fundamentals}, inference problem was investigated both in theory and in practice~\cite{DBLP:journals/pvldb/SzlichtaGGZ13}, which led to the proof of its co-NP-completeness.

Various papers had proposed different mappings that connect those axiomatizations. A paper~\cite{DBLP:journals/pvldb/SzlichtaGGZ13} presents an example of a mapping from LOD to PODs, which indicates that PODs generalize LODs. This generalization is strict, since LODs themselves don't in turn generalize PODs. Authors of~\cite{DBLP:journals/VLDB/SzlichtaGGKS18} propose a polynomial mapping of list-based OD into an equivalent set-based canonical OD, which allows their algorithm to efficiently search for order dependencies. A paper~\cite{DBLP:journals/VLDB/ZijingASS20} claims that PODs strictly generalize canonical ODs, which, combined with the previously mentioned mappings signifies that PODs represent a class of dependencies that is quite generic in nature. An even broader class of dependencies would be Denial Constraints (DCs), which could be found in~\cite{DBLP:journals/VLDB/XuIP13, DBLP:conf/icde/ChuIP13}. PODs are a subset of DCs (LODs, by extension, can also be classified as DCs, since they are a special case of PODs).

Order dependencies can also be generalized via Bidirectional Order Dependencies (BODs), which were analyzed in~\cite{DBLP:journals/VLDB/YifengLZ20, DBLP:journals/VLDB/HemantLI19, DBLP:journals/VPDB/SchmidlP21, DBLP:journals/VLDB/SzlichtaGGKS18}. They allow users to specify the order of sorting for both sides of an OD~\cite{DBLP:journals/pvldb/SzlichtaGGZ13}. Papers~\cite{DBLP:journals/VLDB/SzlichtaGGKS18} and~\cite{DBLP:journals/VLDB/YifengLZ20} had delved into mining those dependencies, while works~\cite{DBLP:journals/VPDB/SchmidlP21} and~\cite{DBLP:journals/VLDB/HemantLI19} had dealt with distributed search. This generalization is primarily useful due to emulating order-by clauses in SQL, which allows for an efficient optimization of such queries~\cite{DBLP:journals/VLDB/YifengLZ20}.

There have also been quite a number of papers researching less generalized dependencies. One example of such dependencies would be Functional Dependencies (FDs), explored in~\cite{DBLP:journals/DMKD/YaoH08,DBLP:journals/VLDB/PapenbrockEMNRSZN15}. According to~\cite{DBLP:journals/vldb/LangerN16}, FDs are a special case of order dependencies. Furthermore, authors of the paper claim that their algorithm ORDER can be used to mine FDs, altough not very efficiently. Better performance can be achieved by algorithms such as FastFD~\cite{DBLP:journals/VLDB/WyssGR01} and Tane~\cite{DBLP:journals/TCJ/HuhtalaKPT99} due to them being designed specifically for mining FDs.

Order Compatibility Dependencies (OCDs), researched in~\cite{DBLP:journals/VLDB/SzlichtaGGKS18, DBLP:journal/ICDE/KaregarMGGKSS22}, are a more specific form of OD. This fact has been put to great use in the work~\cite{DBLP:journals/edbt/ConsonniSMV19} of Cristian Consonni et al. They used the idea of separating ODs into FDs and OCDs to propose a new approach to OD discovery. However, according to the paper~\cite{DBLP:journals/VLDB/YifengLZ20}, this pruning technique can lead to an incomplete set of dependencies being found.

To the best of out knowledge, there exist only two algorithms of exact order dependency discovery. Philipp Langer and Felix Naumann proposed the algorithm ORDER~\cite{DBLP:journals/vldb/LangerN16}, which uses list-based inference rules to traverse list-containment lattice with worst-case time complexity of $O(|\relation{R}|!)$. Jaroslaw Szlichta et al.~\cite{DBLP:journals/pvldb/SzlichtaGGKS17} presented algorithm FASTOD and the set-based inference rules, allowing for faster traversal of set-containment lattice, rather than list-containment one. FASTOD, an improvement of ORDER, is based on a polynomial mapping to a canonical forms of ODs with worst-case time complexity of $O(2^{|\relation{R}|})$. They prove the completeness of their approach and provide scenarios in which ORDER yields incomplete results. Cristian Consonni et al.~\cite{DBLP:journals/edbt/ConsonniSMV19} have made an attempt to improve the existing theory related to set-based ODs, proposing a new algorithm called OCDDISCOVER and showing that it has a significant speedup over FASTOD. Their claims have later been proven to be incorrect~\cite{godfrey2019errata}.

The two major algorithms (FASTOD and ORDER) allow its users to find exact order dependencies adhering to their modern definitions. Implementing these algorithms in an efficient manner and adding a new primitive would allow Desbordante to add yet another concept to its toolkit. On top of that, these algorithms rely on two different axiomatizations, which inspired us to research the differences arising from the distinction in their definitions.

                                                                                 
\section{Algorithms}\label{sec:algos}
\subsection{The General Scheme of Both Algorithms}

FASTOD and ORDER are both lattice-based algorithms for dependency discovery. They initiate the search with either an attribute set or a list (depending on axiomatization) consisting of a single element, and progressively move to larger sets or lists through the lattice, traversing levels one by one. Dependency candidates obtained at a given level are checked for minimality based on the previous levels. Dependencies that pass that check~--- as well as an additional candidate verification check~--- are added to the resulting set of dependencies.

Thus, both FASTOD and ORDER employ a dependency search strategy from small to large. This approach enables the identification of minimal dependencies and efficiently reduces the search space. Partitions may be utilized for a more efficient dependency detection, enabling candidate verification checks in linear time.

Throughout the algorithms' execution, the following stages are repeated, as detailed in the article~\cite{DBLP:journals/pvldb/SzlichtaGGKS17}: dependency search, pruning of the current level, and computation of the next level. This process can be described by the following blocks:

\begin{enumerate}
    \item The first level includes all attributes from the original relation. In the initial iteration, this is the current level.
    \item While the current level is not empty, steps 3-5 are executed.
    \item All dependencies on the current level are identified.
    \item The search space for dependencies is reduced by pruning the current level.
    \item The next level is computed and becomes the current one.
\end{enumerate}

In the following sections, we will describe some of the implementation details that are important for our proposed optimizations. We will also describe details that give a better understanding of the algorithms' approaches to dependency search, with emphasis on pruning in particular.

\subsection{ORDER description}
Algorithm ORDER discovers all \textit{minimal n-ary lexicographical} order dependencies under the operator ``$<$'' (and by Theorem 2.2, all dependencies under ``$\leq$''). 

First, the algorithm determines the columns that can be sorted. These columns are sorted so that sorted partitions can be created according to them, which the algorithm will then work with, without having to access the source data anymore. 

Let's say we sorted some attribute $A$. \textit{SortedPartition}($A$) would then contain equivalence classes that retain the information regarding their indexes prior to sorting. If the values are equal, then their indexes would wind up in the same equivalence class. Sorted partitions are used during validation and allow it to be performed in linear time. It is possible to get sorted partitions for any list of attributes by calculating several products of sorted partitions for single attributes. Sorted partition production is a hash-join-like procedure, which you can learn more about in \cite{DBLP:journals/vldb/LangerN16}.

Next, the algorithm works with a lattice. All lists (permutations) of attributes of length $i$ can be found at the $i$-th level of the lattice. Those permutations supply the algorithm with various candidates, dividing these lists into right and left parts. The algorithm starts from the first level and makes its way down the lattice, increasing value of $i$. Since the algorithm considers lists instead of sets of attributes, there can be a lot of candidates, so an aggressive candidate pruning is applied.

The pruning rules allow the algorithm to immediately establish the validity of a candidate, depending on the candidates that had already been verified. The rules are as follows:
\begin{enumerate}

\item $\attributeList{X} \not\mapsto_{<} \attributeList{Y} \Rightarrow \attributeList{XV} \not\mapsto_{<} \attributeList{Y}$;
\item $\attributeList{X} \mapsto_{<}\attributeList{Y}$ valid $ \Rightarrow \attributeList{X} \mapsto_{<} \attributeList{YW}$ valid;
\item $\attributeList{X} \not\mapsto_{<} \attributeList{Y} \Rightarrow \attributeList{XV} \not\mapsto_{<} \attributeList{YW}$, where $\attributeList{X} \not\mapsto_{<}\attributeList{Y}$ is an invalidation by \textit{swap}. 
\item $\attributeList{X} \mapsto_{<} \attributeList{Y}$ is valid $\Rightarrow\attributeList{XV} \mapsto_{<} \attributeList{YW}$ is valid, if $\attributeList{X}$ contains only unique values.
\end{enumerate}
The attribute lists $\attributeList{X}, \attributeList{Y}, \attributeList{V}, \attributeList{W}$ do not overlap, and only $\attributeList{V}, \attributeList{W}$ can be empty.

\subsection{FASTOD description}

FASTOD is an algorithm for efficient discovery of complete and minimal set of set-based canonical ODs. While ORDER traverses a lattice of all lists of attributes, FASTOD traverses a lattice of all sets of attributes. The idea of the algorithm is to utilize polynomial mapping of order dependencies to canonical forms, which allows it to achieve greater performance: its worst-case time complexity is $O(2^{|\relation{R}|})$.

Instead of regular partitions, FASTOD uses StrippedPartitions, in which equivalence classes with cardinality of 1 are excluded. This sort of compression allows for additional efficiency, but does not interfere with correctness of the algorithm. After calculating the partitions for individual attributes, the FASTOD evaluates partitions for subsequent levels, consisting of several attributes, in linear time using the product of partitions. So, partitions are not calculated from scratch, but are instead derived from previous levels: $\Pi_{\attribute{A} \cup \attribute{B}} = \Pi_{\attribute{A}} * \Pi_{\attribute{B}}$. This dramatically improves the performance of the algorithm.

FASTOD also employs a specific method for storing candidates. They are stored in
$C_c^+(\attributeSet{X}) = \{ \attribute{A} \in \relation{R}: \forall \, \attribute{B} \in \attributeSet{X} \,\, \attributeSet{X} \backslash \{\attribute{A}, \attribute{B}\} : \emptyList \rightarrow_{cst} \attribute{B} \mbox{ does not hold} \}$
and
$C_s^+(\attributeSet{X}) = \{ \{\attribute{A}, \attribute{B}\} \in \attributeSet{X}^2: \attribute{A} \neq \attribute{B} \mbox{ and } \forall \, \attribute{C} \in \attributeSet{X} \, \attributeSet{X} \backslash \{\attribute{A}, \attribute{B}, \attribute{C}\}: \attribute{A} \sim \attribute{B} \mbox{ does not hold, and } \forall \, \attribute{C} \in \attributeSet{X} \, \attributeSet{X} \backslash \{\attribute{A}, \attribute{B}, \attribute{C}\}: \emptyList \rightarrow_{cst} \attribute{C} \mbox{ does not hold } \}$, where $\relation{R}$ denotes the original relation.
This approach prevents candidate sets from growing in size during algorithm's execution. Another benefit of using this representation lies in simplicity of pruning: a set of attributes $X$ is deleted from a level (for all levels above 1) if both sets $C_c^+(\attributeSet{X})$ and $C_s^+(\attributeSet{X})$ are empty.

\subsection{ORDER and FASTOD baselines}

Both baselines are established by reimplementing the corresponding Java algorithm in C++. These adaptations involve minimal changes, addressing the absence of certain Java language features and specific data structures. Furthermore, both C++ implementations incorporate all data types supported by Desbordante, whereas the original FASTOD algorithm was tailored exclusively to integer columns. The C++ implementations abstain from utilizing specialized third-party libraries, such as libraries designed for specific memory management (allocators). Instead, they employ data structures from the standard library and incorporate Boost (https://www.boost.org).

\subsection{FASTOD optimizations}

Internally, the algorithm employs a specialized data structure called stripped partition (further referred as ``partition''), which stores information about the partitioning of the dataset into equivalence classes. The FASTOD algorithm relies on numerous computationally intensive operations involving these partitions (represented by StrippedPartition class in the code).

Through the analysis of real datasets, it has been observed that a considerable portion of them contains columns predominantly composed of blocks of identical values. Some datasets are mainly comprised of such columns. Consequently, a decision was made to optimize the internal representation of partitions.

The standard approach involves storing the indices of all values within each equivalence class. However, this approach proves to be wasteful when encountering a large number of consecutive values falling into the same equivalence class. This results in excessive memory usage and the need for repeated copying of a substantial amount of data. For instance, consider the attribute $I = (1, \ldots, 1, 3, 1, \ldots, 1)$ and the equivalence class representation for the value of 1. In this case, it would appear as follows: $[1] = (0, 1, \ldots, N, N+2, N+3, \ldots, M)$, where $N$ is the index of the 1 located before 3, and $M$ is the index of the last 1.

The \underline{first optimization} involves storing data not as a list of values, but as a list of ranges instead. Revisiting the same attribute $I$, the equivalence class representation for 1 would now be: $[1] = (0 \mbox{--} N, (N+2) \mbox{--} M)$. In this case, it is evident that memory is used much more efficiently, and less data needs to be copied. We call such approach a \textbf{range-based partition representation} (represented by RangeBasedStrippedPartition class in the code).

However, such representation would be inefficient for attributes that do not have a sufficiently large sequences of identical values. In such cases, many small or even degenerate ranges would be observed, not only occupying a significant amount of memory, but also slowing down partition operations.

The \underline{second optimization} addresses this issue. It utilizes knowledge accumulated during dataset preprocessing. At this stage, values in each column are analyzed. If the proportion of values forming ranges is greater than a constant of 0.001, the corresponding attribute is flagged. The optimization involves mindful selection of the representation for the initial partition. If it is constructed based on an attribute flagged as having a sufficiently large proportion of range-forming values, the range-based partition representation is selected. Otherwise, the algorithm uses the standard partition representation. This achieves increased performance on attributes of a specific type without sacrificing performance on other attributes.

Furthermore, it has been observed that the size of value ranges does not increase as a result of partition operations; it usually decreases gradually down to a degenerate range containing a single value. Therefore, starting from a certain point, the representation based on ranges begins to slow down partition operations and expend unnecessary memory. The \underline{third optimization} addresses this issue by dynamically switching the representation from range-based to the standard version. During partition operations, the percentage of small ranges (those with a size less than 40) relative to all ranges in its new state is calculated. If this ratio exceeds a constant of 0.5, the partition representation is switched to a standard one. This achieves the following effect: speed up during the initial stages when dealing with a large portion of range-forming values, and, when the ranges exhaust themselves (mostly turning into small ones) and start slowing down partition operations, the representation is switched to the standard one, retaining the performance characteristics of a baseline approach.

The new representation of partitions allows for performance gains. In particular, this is achieved through a special algorithm for computing range-based partition product. It uses the idea of fast range intersections, which is the basis of partition representation.

For each attribute, a list of ``value-range'' pairs is built, where range means the range of indices in the attribute followed by the corresponding value. By sorting this list by the second element of the pair, we obtain a new list, which we will call $SI$. If we sequentially expand the ranges of each of its pairs, an ordered sequence of indices spanning from zero to the number of table rows minus one is formed. Let's call this sequence $SEQ$. Next, for each such list we create a correspondence table $T$ for each element $a$ in $SEQ$ with an index in $SI$. This index points to a pair whose range contains $a$. This table, together with the ordering condition of the $SI$ list, will allow the algorithm to intersect an arbitrary range $r$ with a list $SI$ of the corresponding attribute in a short time. Instead of sequentially intersecting $r$ with each range in $SI$ (as would be the case if we represented the attribute as an arbitrary list of pairs), we first find two indices in constant time (these indices point to the ranges that contain the beginning and the end of $r$). Then we intersect $r$ with ranges whose indices lie between the two found indices (including the ends). In this case, the ranges obtained as a result of intersection are matched to the same values that correspond to the ranges in $SI$. Combining the resulting ``value-range'' pairs will give us a list that will be the desired result of the intersection of the range $r$ with $SI$.

It is also possible to intersect a list of ranges with $SI$. To do this, you need to intersect each range of this list with $SI$ and combine the results.

Our partition representation involves storing a list of ranges for each equivalence class. This way it can be easily mapped to a list of ``value-range'' pairs by adding a corresponding value for each range. The product of an existing partition with another one, built by some attribute $\attribute{A}$, can be calculated by intersecting this list with a list built by $\attribute{A}$, according to the considerations described earlier.

To better understand the described principle, consider the following example. Let us have an attribute $\attribute{A} = (5, 5, 6, 6, 5, 5, 8)$. The list of ``value-range'' pairs for it will look like $I = \{(5, [0\mbox{--}1]); (5, [4\mbox{--}5]); (6, [2\mbox{--}3]); (8, [6\mbox{--}6])\}$. Sorting it by the second element of the pair (that is, by ranges) gives us $SI_{\attribute{A}} = \{(5, [0\mbox{--}1]); (6, [2\mbox{--}3]); (5, [4\mbox{--}5]); (8, [6\mbox{--}6])\}$. Next, we create a correspondence table: $T_\attribute{A} = \{(0 \rightarrow 0); (1 \rightarrow 0); (2 \rightarrow 1); (3 \rightarrow 1); (4 \rightarrow 2); (5 \rightarrow 2); (6 \rightarrow 3)\}$. Suppose that we have a partition $\Pi_\attributeSet{X}$ and it's range-based representation $\{[0\mbox{--}1], [5\mbox{--}6], [2\mbox{--}4]\}$, where ranges $[0\mbox{--}1], [5\mbox{--}6]$ form the first equivalence class $C_1$, and range $[2\mbox{--}4]$ forms the second one, $C_2$. Lets say we want to calculate $\Pi_{\attributeSet{X} \cup \attribute{A}} = \Pi_{\attributeSet{X}} * \Pi_{\attribute{A}}$, where $\Pi_{\attribute{A}}$ is a partition built for the attribute $\attribute{A}$. To do this, we need to intersect each equivalence class from $\Pi_\attributeSet{X}$ with the list formed by the attribute $\attribute{A}$, that is, with $SI_\attribute{A}$. As an example, consider the intersection of the second equivalence class containing only a single range $d = [2\mbox{--}4]$ with $SI_\attribute{A}$. We calculate the indices of the ranges containing the beginning and the end of $d$: $T_\attribute{A}[2] = 1$, $T_\attribute{A}[4] = 2$. In the list $SI_\attribute{A}$, there are two ranges between indices 1 and 2: $[2\mbox{--}3]$ and $[4\mbox{--}5]$. We intersect them with $d$, which gives us $[2\mbox{--}4] \cap [2\mbox{--}3] = [2\mbox{--}3]$, $[2\mbox{--}4] \cap [4\mbox{--}5] = [4\mbox{--}4]$. In this case, the resulting ranges matched to the same values that corresponded to the ranges in $SI_\attribute{A}$. So $M = C_2 \cap SI_\attribute{A} = \{ (6, [2\mbox{--}3]); (5, [4\mbox{--}4]) \}$. If the equivalence class consisted of several ranges, we would intersect each of them with $SI_\attribute{A}$, and then combine the results into a single list. Now we sort $M$ by the first element of the pair and extract equivalence classes from it. This gives us two equivalence classes, the representation of which using indexes is as follows: $[4\mbox{--}4]$ and $[2\mbox{--}3]$. We exclude the equivalence classes with cardinality of 1, so only $[2\mbox{--}3]$ is added to the resulting list. After performing similar operations with each equivalence class from $\Pi_\attributeSet{X}$, we obtain the final list of equivalence classes, which will be the result of the product of partitions.

A description of the algorithm in a more general form is presented in the Algorithm~\ref{alg:rb_partition_product}.

\begin{algorithm}
    \caption{Range-based partition product}\label{alg:rb_partition_product}
    \KwData{$T_\attribute{A}$~--- correspondence table for attribute \attribute{A}, $\Pi_\attributeSet{X}$~--- first partition, $SI_\attribute{Y}$~--- sorted value-range representation of second partition $\Pi_{\attribute{A}}$ of $\attribute{A}$}
    \KwResult{$O$~--- partition $\Pi_{\attributeSet{X} \cup \attribute{A}}$}
    
    $O \gets \emptyset$\;
    \For{$C \in \Pi_\attributeSet{X}$} {
        $M \gets \emptyset$\;
        \For{$[d_s; d_e] \in C$} {
            $s \gets T_\attribute{A}[d_s]$\;
            $e \gets T_\attribute{A}[d_e]$\;
            $R \gets SI_\attribute{A}[s \ldots e]$\;
        
            \For{$(v, [r_s; r_e]) \in R$} {
                $r \gets [d_s; d_e] \cap [r_s; r_e]$\;
                $\mbox{add } (v, r) \mbox{ to } M$\;
            }
        }
        $\mbox{sort } M$\;
        $E \gets \mbox{extract equivalence classes from } M$\;
        $\mbox{exclude degenerate classes from } E$\;
        $\mbox{add each element from } E \mbox{ to } O$
    }
    
    $\mbox{return } O$\;
\end{algorithm}

\subsection{ORDER optimizations}

Unlike FASTOD, optimizations for ORDER are based on using more efficient data structures and sorting algorithms from the Boost library where it is most needed.

After studying the performance of the algorithm, we concluded that in the vast majority of cases, the following operations take the most time: sorting attribute values when creating sorted partitions, searching for elements of equivalence classes during validity checks, and calculating the product of sorted partitions.

\textbf{Sorting.} Sorting of values occurs at the start of the algorithm in order to obtain efficient data representation~--- sorted partitions. As a result, the source data is not used on the next steps of the algorithm. 

This step is the primary target for optimization, since it will result in performance gains on all datasets, unlike calculation of the product of sorted partitions, which may not occur due to dependencies not necessarily being found.

In addition to the sorts offered in the C++ standard library, we have the opportunity to use Boost.Sort library, which offers a set of different sorts, both parallel and serial. For single-threaded execution, we select flat\_stable\_sort because it offers decent performance and low memory consumption. block\_indirect\_sort was chosen for multi-threaded execution, for the same reasons. Replacing the sorting algorithm from the standard C++ library can bring improvements in performance and memory consumption.

\textbf{Candidate validation}. Validation is performed using a pair of sorted partitions. Algorithm goes through pairs of equivalence classes, in which identical elements are searched. The most efficient structure for a large number of searches is unordered\_set, which has several implementations. In addition to the implementation from the C++ standard library, Boost offers its own implementation: unordered\_flat\_set, which has high performance as its main characteristic. Using unordered\_flat\_set can theoretically increase the performance of the algorithm.

\textbf{Calculation of Sorted Partitions Product}. Partition product is calculated for datasets that have dependencies, and the more dependencies there are in the dataset, the more often that product occurs. Product is calculated using hash maps, so using a more efficient unordered\_flat\_map from the Boost instead of unordered\_map from the standard C++ library can bring an increase in speed to datasets that contain a large number of dependencies.

                                                                                         
\section{Experiments and Discussion}\label{sec:exper}
\subsection{General}

To evaluate our techniques, we have developed our own implementations of both algorithms~--- FASTOD and ORDER~--- and experimentally compared them with the existing implementations written in Java. We present research questions and report experimental results for each algorithm in the next sections. Note that we do not compare FASTOD and ORDER with each other, as it is meaningless since they are designed to yield different results.

In our experiments, we have only considered run time of the algorithm itself, as parsing and preprocessing differ in Desbordante and Java implementations, and therefore could skew the final results. We leave the parsing and data preprocessing stages for future investigations.

Each discussed experiment was repeated three times, and the average of the results was calculated.

\textbf{Experimental Setup.}
Experiments were performed using the following hardware and software configuration. Hardware: Intel® Core™ i7-11800H CPU @ 2.30GHz (8 cores), 16GB DDR4 3200MHz RAM, 512GB SSD SAMSUNG MZVL2512HCJQ-00BL2. Software: Kubuntu 23.10, Kernel 6.5.0-14-generic (64-bit), gcc 13.2.0, openjdk 11.0.21 2023-10-17, OpenJDK Runtime Environment (build 11.0.21+9-post-Ubuntu-0ubuntu123.10), OpenJDK 64-Bit Server VM (build 11.0.21+9-post-Ubuntu-0ubuntu123.10, mixed mode, sharing).

\subsection{FASTOD}

\textbf{Methodology.}
The Java implementation of the FASTOD algorithm (https://github.com/leveretconey/cocoa/tree/master/ src/main/java/leveretconey/fastod), unlike its C++ counterpart, is limited to datasets composed solely of integer values. Consequently, the original datasets were transformed to adhere to the specified format before running experiments. The resulting datasets can be found in the corresponding repository (https://github.com/Sched71/Desbordante-OD-Data).

For FASTOD we pose the following research questions:
\begin{enumerate}

    \item[RQ1] Is it possible to outperform existing implementation by simply reimplementing OD discovery algorithm in C++?

    \item[RQ2] What improvement does the proposed range-based partition representation offer on datasets with columns containing ranges?
    
    \item[RQ3] What is the overhead of the proposed range-based partition representation? It is true that using this representation does not compromise algorithm performance on regular datasets?

    \item[RQ4] How well does the performance of the C++ implementations scale with the increase in the number of columns?
    
    \item[RQ5] What are the memory savings of both C++ implementations?    
\end{enumerate}

To answer these questions, we have designed the following experiments for each RQ:

\begin{enumerate}
    \item In the first experiment, we are comparing the vanilla C++ implementation of the FASTOD algorithm with its counterpart written in Java.
    
    \item In the second experiment, we are comparing the baseline C++ implementation with the one containing the proposed technique~--- the range-based partition representation. 
    
    \item In the third experiment, we are comparing the same approaches as in the second experiment, but on datasets containing little range data. Our goal is to demonstrate that this optimization does not compromise performance on typical datasets (i.e., those that contain little range data).

    \item In the fourth experiment, we study how well the C++ implementations scale with the number of columns in the dataset.
    
    \item The fifth experiment evaluates memory savings for both C++ implementations compared to the Java implementation.
\end{enumerate}

\textbf{Evaluation.}
To conduct experiments, we used the datasets shown in Table~\ref{tab:dataset_description_fastod}. It includes a description of the datasets, as well as their short names, which we will use for brevity. The overall results are shown in Table~\ref{tab:overall_results_fastod}. We grayed out rows with special datasets containing columns with a large number of ranges.

\textbf{Experiment 1.}
In this experiment, we compare our baseline implementation of the FASTOD algorithm with the Java implementation. The results of the conducted experiments are presented in Table~\ref{tab:overall_results_fastod}. The experiments unequivocally demonstrated that the C++ implementation of the algorithm exhibits higher performance compared to the Java implementation. We outperform it by a factor of up to 8, with 4 being the average.

\textbf{Experiment 2.}
Our second experiment demonstrates possible performance increase on special datasets that contain numerous repeated values in the columns. The results are presented in Table~\ref{tab:overall_results_fastod}, where we have highlighted in gray the rows with special datasets. The experiments demonstrated that the new partition representation can contribute to achieving a speedup of 1.3x--1.8x.

\textbf{Experiment 3.}
This experiment is similar to the first one, except the comparison is now conducted between the baseline and optimized versions of the C++ Desbordante implementation of the algorithm. The results of the conducted experiments are presented in Table~\ref{tab:overall_results_fastod}. As evident from the results, not only did the algorithm's runtime not increase on typical datasets after the application of optimizations, but it even showed a slight decrease. Thus, it can be concluded that our optimizations not only do not deteriorate the original implementation, but also enable significant improvements in processing time on datasets of a specific nature.

\textbf{Experiment 4.}
In this experiment, we studied the dependence of the algorithm's running time on the number of columns in the corresponding dataset. We used typical dataset A and special dataset D3 containing many columns with ranges as our initial datasets. We excluded a certain number of columns from each dataset, starting from the beginning of the dataset, thereby obtaining a dataset with the required number of columns.

\begin{figure}[h]
    \begin{center}
        \frame{\includegraphics[width=0.9\linewidth]{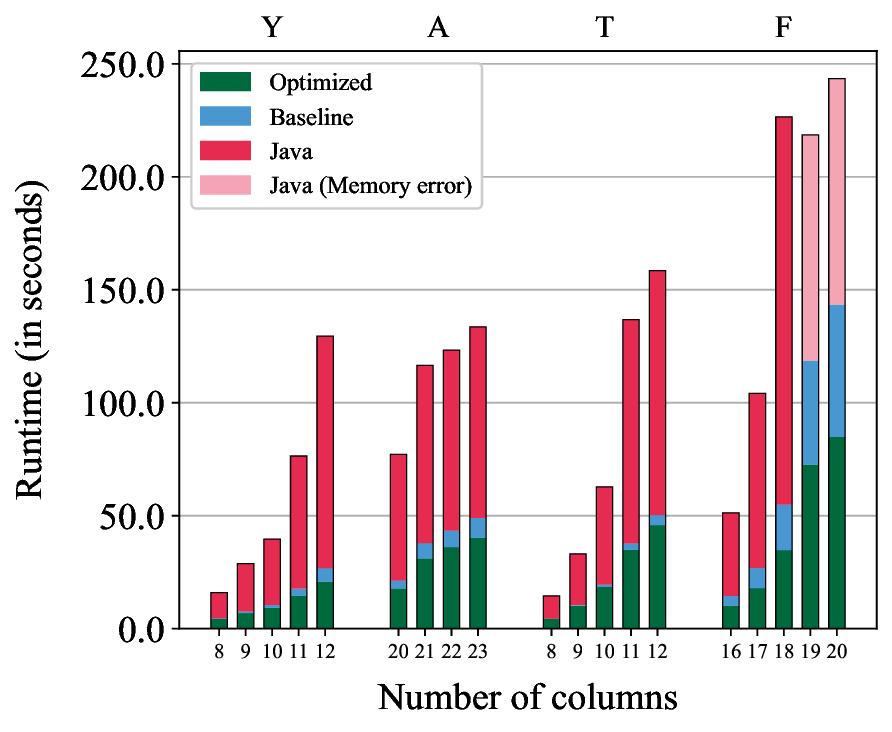}}
    \end{center}
    \caption{Scalability in number of columns, part 1 (FASTOD)}
    \label{plot:diabetes_norm}
\end{figure}

\begin{figure}[h]
    \begin{center}
        \frame{\includegraphics[width=0.9\linewidth]{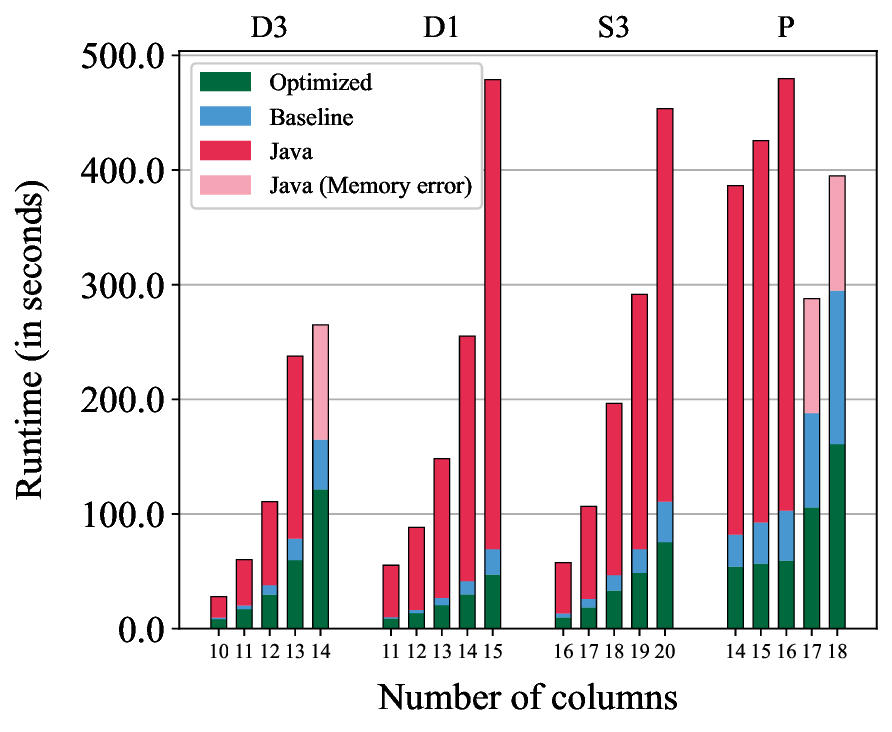}}
    \end{center}
    \caption{Scalability in number of columns, part 2 (FASTOD)}
    \label{plot:anonymize_norm}
\end{figure}

The results of this experiment are demonstrated in Figures~\ref{plot:diabetes_norm} and~\ref{plot:anonymize_norm}. They both show a clear superiority of our implementations compared to the Java version, as well as the speedup due to a special representation of partitions, which is clearly visible in Figure~\ref{plot:diabetes_norm}.

\noindent
\noindent

\begin{figure}
    \begin{center}
        \begin{tikzpicture}[framed, scale=0.9]
            \begin{axis}[
                ylabel=Memory usage (in gigabytes),
                xbar,
                ybar,
                ymajorgrids,
                ymin=0, ymax=12.5,
                xmin=0, xmax=5,
                xtick={1,2,3,4},
                xticklabels={G,A,S2,S1},
            	legend style={
                    at={(0.5,1.15)},
            	    anchor=north,
                    legend columns=-1
                },
            ]

            \addplot[color=black, fill=dartmouthgreen, opacity=1] 
                coordinates {
                (1, 1.160) (2, 1.270) (3, 0.266) (4, 4.504)
                };
            
            \addplot[color=black, fill=celestialblue, opacity=1] 
            	coordinates {
                (1, 0.809) (2, 1.035) (3, 0.261) (4, 3.478)
                };

            \addplot[color=black, fill=amaranth, opacity=1]
                coordinates {
                (1, 3.150) (2, 3.087) (3, 0.762) (4, 11.919)
                };
                
            \legend{Optimized,Baseline,Java}
            
            \end{axis}
        \end{tikzpicture}
    \end{center}
    \caption{Memory usage (FASTOD)}
    \label{fig:memory_usage_fastod}
\end{figure}
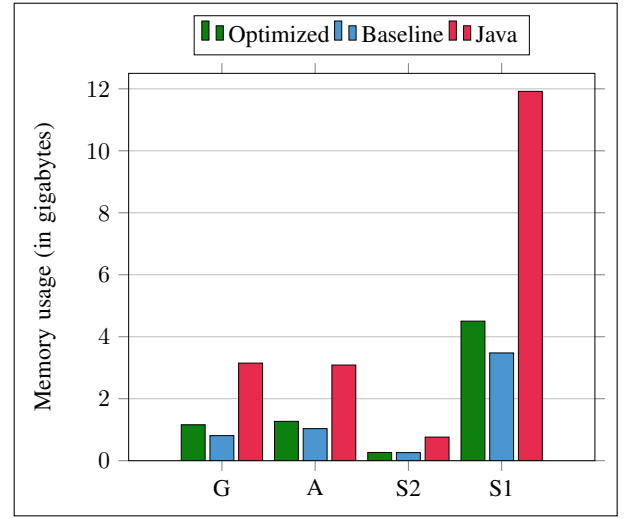

\textbf{Experiment 5.}
Our last experiment shows memory usage of all FASTOD implementations. The testing involved datasets G, A, S1, S2 and its results are presented in Figure~\ref{fig:memory_usage_fastod}.

We can observe a significant reduction in RAM consumption by both of our implementations of the algorithm. Baseline C++ implementation outperforms Java implementations by 2.9--3.9 times and the optimized implementation outperforms it by 2.4--2.9 times. It is also noticeable that the optimized implementation consumes more memory than the baseline one. This is a necessary price to pay for reducing execution time.

In addition, we found that the Java implementation consumes too much memory. For example, its execution on D3~--- which includes 14 columns~--- ends with a memory error: there is not enough RAM on the computer on which the test was executed. This is demonstrated in Figure~\ref{plot:diabetes_norm}, where semi-transparent bar represents the memory error. The same situation is observed on some other datasets. We reflected this in the Table~\ref{tab:overall_results_fastod} using the same notation. At the same time, our implementations successfully tackle all these tasks.

\subsection{ORDER}

\textbf{Methodology.} For the ORDER algorithm, we pose the following research questions:
\begin{enumerate}

    \item[RQ1] Is it possible to outperform existing implementation by simply re-implementing OD discovery algorithm in C++?

    \item[RQ2] Which of the proposed optimizations can improve performance, and will their simultaneous application be effective?

    \item[RQ3] Can an optimized C++ implementation provide memory savings?
\end{enumerate}

To answer these questions, we have designed the following experiments for each RQ:

\begin{enumerate}
    \item In the first experiment, we are comparing the vanilla C++ implementation of the ORDER algorithm with its counterpart written in Java.

    \item In the second experiment, we consider the optimization approaches both individually and simultaneously. We also make a comparison with the base version of the implementation.

    \item In the third experiment, we are comparing memory usage of the optimized C++ implementation of the ORDER algorithm and implementation from Metanome.
\end{enumerate}

\textbf{Evaluation.}
To conduct experiments, we used the datasets
shown in Table~\ref{tab:order_datasets}. It includes a description of the datasets, as well as their short names, which we will use for brevity. The overall results are shown in Table~\ref{tab:order_overall}. We had to select other datasets since ORDER algorithm is much faster than FASTOD. This results in sub second run times, which is not suitable for experiments, as the overall run time is a subject to measurement errors. Note that this does not mean that FASTOD is useless~--- ORDER misses some of the dependencies due to excessive pruning. On the other hand, ORDER can be useful for quick profiling aimed at obtaining a rough picture rather than striving for completeness of the set of discovered dependencies.

\textbf{Experiment 1.} 
In this experiment, we compared the base version of the C++ implementation with the Java implementation in Metanome.

The results of the conducted experiments are presented in Table~\ref{tab:order_overall} in columns Java, Base and Impr (base).

Experiments have shown that the base implementation in C++ is superior to the Java implementation in most cases. We outperform it by a factor of up to 9, with 4 being the average. 

Due to the presence of datasets on which Java has higher performance, it became clear that additional optimizations were necessary.

\textbf{Experiment 2.}
In this experiment, we compared the base C++ implementation with implementations where the proposed optimizations were applied. Optimizations were applied both individually and simultaneously.

Overall, we have compared boost::unordered\_flat\_map (flat\_map), boost::unordered\_flat\_set (flat\_set), and boost::block\_indirect\_sort (sort) to their standard counterparts~--- std::unordered\_map, std::unordered\_set, and std::sort. The resulting ratio is presented in Table~\ref{tab:order_results_baseline_optimized}. The last column contains the results of three optimizations combined. 

\begin{table}[htp]
    \centering
    \caption{C++ optimized improvements in comparison to base (ORDER)}
    \label{tab:order_results_baseline_optimized}    
    \begin{tabular}{|c|c|c|c|c|}
        \hline
        Dataset & flat\_map & flat\_set & sort & Final\\
        \hline \hline
        \rowcolor{LightGray}
        Diabetes & 1.017x & 4.954x & 0.969x & 6.253x\\
        \hline
        Pfw & 1.019x & 1.850x & 1.015x & 2.258x\\
        \hline
        \rowcolor{LightGray}
        Ditag & 1.026x & 1.267x & 1.546x & 2.189x\\
        \hline
        Credit & 1.003x & 1.128x & 1.421x & 1.727x\\
        \hline
        \rowcolor{LightGray}
        Epic & 1.023x & 1.480x & 1.268x & 2.128x\\
        \hline
        Modis & 1.025x & 1.592x & 1.341x & 2.550x\\
        \hline
        \rowcolor{LightGray}
        Bay & 1.026x & 1.360x & 1.503x & 2.785x\\
        \hline
    \end{tabular}
\end{table}

According to the results of the experiments, it can be concluded that the use of unordered\_flat\_map did not bring significant improvement, unlike the use of sort from Boost and the use of unordered\_flat\_set. In addition, applying all optimizations at the same time gives even bigger performance boost than the product of the performance increases obtained by testing individual optimizations.

\textbf{Experiment 3.}
In the third experiment, we compared memory usage of the optimized C++ implementation with the Metanome version. The results of the conducted experiments are presented in Table~\ref{tab:order_memory}.

\begin{table}[htp]
    \centering
    \caption{Metanome vs Desbordante memory consumption (ORDER)}
    \label{tab:order_memory}    
    \begin{tabular}{|c|c|c|c|}
        \hline
Dataset  & C++ (MB) & Java (MB) & \multicolumn{1}{l|}{Improvement} \\ \hline \hline
\rowcolor{LightGray}
Diabetes & 177.94   & 429.37    & 2.413x                           \\ \hline
Pfw      & 150.69   & 265.62    & 1.762x                           \\ \hline
\rowcolor{LightGray}
Ditag    & 3715.66  & 5951.69   & 1.601x                           \\ \hline
Credit   & 5607.87  & 7333.7    & 1.307x                           \\ \hline
\rowcolor{LightGray}
Epic     & 662.78   & 1309.7    & 1.976x                           \\ \hline
Modis    & 2614.75  & 6240.43   & 2.386x                           \\ \hline
\rowcolor{LightGray}
Bay      & 5690.77  & 7345.08   & 1.290x                           \\ \hline
    \end{tabular}
\end{table}

Experiments have shown that our implementation uses less memory compared to the implementation from Metanome. To be specific, we achieved memory savings of 1.3--2.4x, depending on the dataset.


\begin{table*}[htp]
    \centering
    \caption{Dataset Description (FASTOD)}
    \label{tab:dataset_description_fastod}    
    \begin{tabular}{|c|c|c|c|c|c|c|c|}
        \hline
        Dataset & Short name & Columns  & Rows & Size (MB) & \#OD & \#FD & \#OCD \\
        \hline \hline
        \rowcolor{LightGray}
        graduation\_dataset\_norm\_15c.csv & G & 15 & 4424 & 0.14 & 333 & 1 & 332 \\
        \hline
        Anonymize\_norm.csv & A & 23 & 38480 & 4.28 & 115400 & 7774 & 107626 \\
        \hline
        \rowcolor{LightGray}
        PFW\_2021\_public\_norm.csv & P & 18 & 100000 & 7.51 & 1240 & 48 & 1192 \\
        \hline
        Spotify\_Dataset\_V3\_norm.csv & S1 & 11 & 651936 & 35.96 & 1645 & 171 & 1474 \\
        \hline
        \rowcolor{LightGray}
        diabetes\_binary\_norm.csv & D3 & 14 & 67136 & 2.18 & 0 & 0 & 0 \\
        \hline
        spotify-2023\_norm.csv & S2 & 20 & 953 & 0.06 & 198327 & 17915 & 180412 \\
        \hline
        \rowcolor{LightGray}
        diabetes\_binary2\_norm.csv & D4 & 12 & 236378 & 6.99 & 0 & 0 & 0 \\
        \hline
        Dataset\_norm.csv & D1 & 15 & 175028 & 13.64 & 858 & 68 & 790 \\
        \hline
        \rowcolor{LightGray}
        file.csv & F & 20 & 52955 & 7.08 & 3134 & 70 & 3064 \\
        \hline
        DOSE\_V2\_norm.csv & D2 & 16 & 46797 & 5.18 & 5912 & 592 & 5320 \\
        \hline
        \rowcolor{LightGray}
        merged\_data\_norm.csv & M & 4 & 11509051 & 276.06 & 0 & 0 & 0 \\
        \hline
        Test\_norm.csv & T & 12 & 89786 & 3.17 & 326 & 12 & 314 \\
        \hline
        \rowcolor{LightGray}
        openpowerlifting\_norm.csv & O & 13 & 386414 & 33.50 & 1079 & 19 & 1060 \\
        \hline
        youtube\_norm.csv & Y & 12 & 161470 & 13.35 & 1752 & 96 & 1656 \\
        \hline
        \rowcolor{LightGray}
        superstore\_norm.csv & S3 & 20 & 51290 & 6.47 & 45916 & 2098 & 43818 \\
        \hline
    \end{tabular}
\end{table*}

\begin{table*}[htp]
    \centering
    \caption{Overall Results (FASTOD)}
    \label{tab:overall_results_fastod}    
    \begin{tabular}{|c|c|c|c|c|c|c|c|c|}
        \hline
        \multirow{2}{*}{Dataset} & \multirow{2}{*}{\#Columns} & \multirow{2}{*}{\#RB-columns} & Java & Base & Optimized & \multirow{2}{*}{Impr (base)} & \multirow{2}{*}{Impr (opt)} & \multirow{2}{*}{Impr (total)} \\
        & & & (seconds) & (seconds) & (seconds) & & & \\
        \hline \hline
        \rowcolor{LightGray}
        G & 15 & 8 & 50.326 & 17.083 & 14.184 & 2.946x & 1.204x & 3.548x \\
        \hline
        A & 23 & 3 & 135.004 & 49.309 & 40.137 & 2.738x & 1.229x & 3.364x \\
        \hline
        \rowcolor{LightGray}
        P & 18 & 8 & $ME$ & 294.900 & 160.967 & $\infty$ & 1.832x & $\infty$ \\
        \hline
        S1 & 11 & 2 & 724.101 & 93.216 & 71.644 & 7.768x & 1.301x & 10.107x \\
        \hline
        \rowcolor{LightGray}
        D3 & 14 & 11 & $ME$ & 166.503 & 121.866 & $\infty$ & 1.366x & $\infty$ \\
        \hline
        S2 & 20 & 2 & 7.177 & 5.187 & 4.994 & 1.384x & 1.039x & 1.437x \\
        \hline
        \rowcolor{LightGray}
        D4 & 12 & 8 & $ME$ & 189.312 & 142.254 & $\infty$ & 1.331x & $\infty$ \\
        \hline
        D1 & 15 & 3 & 478.885 & 69.236 & 47.129 & 6.917x & 1.469x & 10.161x \\
        \hline
        \rowcolor{LightGray}
        F & 20 & 15 & $ME$ & 143.485 & 84.933 & $\infty$ & 1.689x & $\infty$ \\
        \hline
        D2 & 16 & 2 & 23.348 & 7.176 & 6.676 & 3.254x & 1.075x & 3.497x \\
        \hline
        \rowcolor{LightGray}
        M & 4 & 4 & 49.791 & 16.723 & 10.438 & 2.977x & 1.602x & 4.770x \\
        \hline
        T & 12 & 6 & 158.503 & 50.214 & 45.864 & 3.157x & 1.095x & 3.456x \\
        \hline
        \rowcolor{LightGray}
        O & 13 & 7 & $ME$ & 264.255 & 182.810 & $\infty$ & 1.446x & $\infty$ \\
        \hline
        Y & 12 & 5 & 129.516 & 26.755 & 20.810 & 4.841x & 1.287x & 6.224x \\
        \hline
        \rowcolor{LightGray}
        S3 & 20 & 10 & 453.583 & 110.795 & 75.537 & 4.094x & 1.467x & 6.005x \\
        \hline
    \end{tabular}
\end{table*}

\begin{table*}[htp]
    \centering
    \caption{Overall Memory Usage Results (FASTOD)}
    \label{tab:overall_memory_usage_results_fastod} 
    \begin{tabular}{|c|c|c|c|c|c|}
    \hline
    Dataset  & Java (GB) & Base (GB) & Optimized (GB) & Java vs Base & Java vs Optimized \\
    \hline \hline
    \rowcolor{LightGray}
    G & 3.150 & 0.809 & 1.160 & 3.894x & 2.716x \\
    \hline
    A & 3.087 & 1.035 & 1.270 & 2.983x & 2.431x \\
    \hline
    \rowcolor{LightGray}
    P & $ME$ & 9.321 & 14.304 & $\infty$ & $\infty$ \\
    \hline
    S1 & 11.919 & 3.478 & 4.504 & 3.427x & 2.646x \\
    \hline
    \rowcolor{LightGray}
    D3 & $ME$ & 8.296 & 14.124 & $\infty$ & $\infty$ \\
    \hline
    S2 & 0.762 & 0.261 & 0.266 & 2.920x & 2.865x \\
    \hline
    \rowcolor{LightGray}
    D4 & $ME$ & 7.233 & 12.062 & $\infty$ & $\infty$ \\
    \hline
    D1 & 6.878 & 2.374 & 3.556 & 2.897x & 1.934x \\
    \hline
    \rowcolor{LightGray}
    F & $ME$ & 6.031 & 8.941 & $\infty$ & $\infty$ \\
    \hline
    D2 & 1.062 & 0.071 & 0.126 & 14.958x & 8.423x \\
    \hline
    \rowcolor{LightGray}
    M & 9.508 & 2.519 & 2.595 & 3.775x & 3.664x \\
    \hline
    T & 2.537 & 0.760 & 0.994 & 3.338x & 2.552x \\
    \hline
    \rowcolor{LightGray}
    O & $ME$ & 8.109 & 11.215 & $\infty$ & $\infty$ \\
    \hline
    Y & 3.168 & 0.839 & 1.248 & 3.776x & 2.538x \\
    \hline
    \rowcolor{LightGray}
    S3 & 7.142 & 2.698 & 3.915 & 2.647x & 1.824x \\
    \hline
    \end{tabular}
\end{table*}

\begin{table*}[htp]
\centering
\caption{Dataset Description (ORDER)}
\label{tab:order_datasets} 
\begin{tabular}{|c|c|c|c|c|c|}
\hline
Dataset                                             & Short name & Columns & Rows    & Size (MB) & \#OD \\ \hline
\hline
\rowcolor{LightGray}
diabetes\_binary\_BRFSS2021.csv & Diabetes   & 22      & 236379  & 17.0      & 0    \\ \hline
PFW\_2021\_public.csv                               & Pfw        & 22      & 100001  & 14.7      & 17     \\ \hline
\rowcolor{LightGray}
DITAG.csv                                           & Ditag      & 5       & 4339917 & 299.0     & 0     \\ \hline
creditcard\_2023.csv                                & Credit     & 31      & 568631  & 324.8     & 0     \\ \hline
\rowcolor{LightGray}
EpicMeds.csv                                        & Epic       & 10      & 1281732 & 56.8      & 9     \\ \hline
modis\_2000-2019\_Australia.csv                     & Modis      & 15      & 5081220 & 410.4     & 7     \\ \hline
\rowcolor{LightGray}
bay\_wheels\_data\_wrangled.csv                     & Bay        & 8       & 5022834 & 660.1     & 0     \\ \hline
\end{tabular}
\end{table*}

\begin{table*}[htp]
\centering
\caption{Overall Results (ORDER)}
\label{tab:order_overall} 
\begin{tabular}{|c|c|c|c|c|c|c|}
\hline
\multirow{2}{*}{Dataset}  & Java  & Base  & Optimized & \multirow{2}{*}{Impr (base)} & \multirow{2}{*}{Impr (opt)} & \multirow{2}{*}{Impr (total)} \\ 
& (seconds) & (seconds) & (seconds) & & & \\
\hline \hline
\rowcolor{LightGray}
Diabetes & 3.331  & 3.696  & 0.591       & 0.901x                           & 6.253x                          & 5.636x                            \\ \hline
Pfw      & 1.361  & 0.472   & 0.209       & 2.883x                           & 2.258x                          & 6.511x                            \\ \hline
\rowcolor{LightGray}
Ditag    & 16.833 & 7.219  & 3.297      & 2.331x                           & 2.189x                          & 5.105x                            \\ \hline
Credit   & 32.585 & 5.279  & 3.056      & 6.172x                           & 1.727x                          & 10.662x                           \\ \hline
\rowcolor{LightGray}
Epic     & 8.649  & 3.438  & 1.615      & 2.515x                           & 2.128x                          & 5.355x                            \\ \hline
Modis    & 88.069 & 9.339  & 3.661      & 9.430x                           & 2.550x                          & 24.055x                           \\ \hline
\rowcolor{LightGray}
Bay      & 33.158 & 18.466 & 6.630      & 1.795x                           & 2.785x                          & 5.001x                            \\ \hline
\end{tabular}
\end{table*}

                                                                                  
\section{Conclusion}\label{sec:concl}
In this paper, we have presented optimization techniques for two state-of-the-art OD discovery algorithms (ORDER and FASTOD), which allow us to add a new important primitive to Desbordante. Our experiments show a significant increase in algorithm performance (up to 10x), as well as a decrease in memory consumption (up to 2.9x). Described optimizations and the new representation of partitions can help optimize any other algorithms with a similar structure of components, which once again signifies the importance of our research.

Both C++ implementations will be of use for Desbordante's end-users, despite aiming at the same problem, namely discovery of ODs. ORDER is significantly faster than FASTOD, but misses some of the dependencies due to excessive pruning. ORDER, on the other hand, can be useful for quick profiling aimed at obtaining a rough picture rather than striving for completeness of the set of discovered dependencies. Finally, both implementations~--- ORDER and FASTOD~--- are open-source (https://github.com/Mstrutov/Desbordante/) and are merged (PRs 294, 355) into the Desbordante.

\section*{Acknowledgment}
We would like to thank Vladislav Makeev for his help with the preparation of the paper.

\bibliographystyle{IEEEtran}

\bibliography{FRUCTexample-short}

\begin{thebibliography}{10}
\providecommand{\url}[1]{#1}
\csname url@samestyle\endcsname
\providecommand{\newblock}{\relax}
\providecommand{\bibinfo}[2]{#2}
\providecommand{\BIBentrySTDinterwordspacing}{\spaceskip=0pt\relax}
\providecommand{\BIBentryALTinterwordstretchfactor}{4}
\providecommand{\BIBentryALTinterwordspacing}{\spaceskip=\fontdimen2\font plus
\BIBentryALTinterwordstretchfactor\fontdimen3\font minus
  \fontdimen4\font\relax}
\providecommand{\BIBforeignlanguage}[2]{{%
\expandafter\ifx\csname l@#1\endcsname\relax
\typeout{** WARNING: IEEEtran.bst: No hyphenation pattern has been}%
\typeout{** loaded for the language `#1'. Using the pattern for}%
\typeout{** the default language instead.}%
\else
\language=\csname l@#1\endcsname
\fi
#2}}
\providecommand{\BIBdecl}{\relax}
\BIBdecl

\bibitem{DBLP:conf/icde/ChuIP13}
X.~Chu, I.~F. Ilyas, and P.~Papotti, ``Holistic data cleaning: Putting
  violations into context,'' in \emph{{ICDE'13}}, C.~S. Jensen \emph{et~al.},
  Eds.\hskip 1em plus 0.5em minus 0.4em\relax {IEEE} Computer Society, 2013,
  pp. 458--469.

\bibitem{10.5555/3312004}
Z.~Abedjan, L.~Golab, F.~Naumann, and T.~Papenbrock, \emph{Data
  Profiling}.\hskip 1em plus 0.5em minus 0.4em\relax Morgan \& Claypool
  Publishers, 2018.

\bibitem{DBLP:journals/corr/abs-2301-05965}
G.~Chernishev \emph{et~al.}, ``Desbordante: from benchmarking suite to
  high-performance science-intensive data profiler,'' \emph{CoRR}, vol.
  abs/2301.05965, 2023.

\bibitem{DBLP:journals/VLDB/PapenbrockEMNRSZN15}
T.~Papenbrock \emph{et~al.}, ``Functional dependency discovery: an experimental
  evaluation of seven algorithms,'' \emph{Proc. VLDB Endow.}, vol.~8, no.~10,
  p. 1082–1093, jun 2015.

\bibitem{10.1145/3357384.3357916}
F.~D\"{u}rsch \emph{et~al.}, ``Inclusion dependency discovery: An experimental
  evaluation of thirteen algorithms,'' in \emph{CIKM'19}, 2019, p. 219–228.

\bibitem{10.5555/2677098}
C.~C. Aggarwal and J.~Han, \emph{Frequent Pattern Mining}.\hskip 1em plus 0.5em
  minus 0.4em\relax Springer Publishing Company, Incorporated, 2014.

\bibitem{10.5555/1315451.1315509}
P.~G. Brown and P.~J. Hass, ``Bhunt: Automatic discovery of fuzzy algebraic
  constraints in relational data,'' in \emph{VLDB'03}.\hskip 1em plus 0.5em
  minus 0.4em\relax VLDB Endowment, 2003, p. 668–679.

\bibitem{10.1145/3292500.3330993}
M.~Hulsebos \emph{et~al.}, ``Sherlock: A deep learning approach to semantic
  data type detection,'' in \emph{SIGKDD'19}, 2019, p. 1500–1508.

\bibitem{DBLP:journals/pvldb/SzlichtaGGKS17}
\BIBentryALTinterwordspacing
J.~Szlichta, P.~Godfrey, L.~Golab, M.~Kargar, and D.~Srivastava, ``Effective
  and complete discovery of order dependencies via set-based axiomatization,''
  \emph{Proc. {VLDB} Endow.}, vol.~10, no.~7, pp. 721--732, 2017. [Online].
  Available: \url{http://www.vldb.org/pvldb/vol10/p721-szlichta.pdf}
\BIBentrySTDinterwordspacing

\bibitem{DBLP:journals/pvldb/SzlichtaGGZ13}
\BIBentryALTinterwordspacing
J.~Szlichta, P.~Godfrey, J.~Gryz, and C.~Zuzarte, ``Expressiveness and
  complexity of order dependencies,'' \emph{Proc. {VLDB} Endow.}, vol.~6,
  no.~14, pp. 1858--1869, 2013. [Online]. Available:
  \url{http://www.vldb.org/pvldb/vol6/p1858-szlichta.pdf}
\BIBentrySTDinterwordspacing

\bibitem{DBLP:journals/TCS/GinsburgH83}
\BIBentryALTinterwordspacing
S.~Ginsburg and R.~Hull, ``Order dependency in the relational model,''
  \emph{Theoretical Computer Science}, vol.~26, no.~1, pp. 149--195, 1983.
  [Online]. Available:
  \url{https://www.sciencedirect.com/science/article/pii/0304397583900841}
\BIBentrySTDinterwordspacing

\bibitem{DBLP:journals/vldb/LangerN16}
\BIBentryALTinterwordspacing
P.~Langer and F.~Naumann, ``Efficient order dependency detection,''
  \emph{{VLDB} J.}, vol.~25, no.~2, pp. 223--241, 2016. [Online]. Available:
  \url{https://doi.org/10.1007/s00778-015-0412-3}
\BIBentrySTDinterwordspacing

\bibitem{9435469}
M.~Strutovskiy, N.~Bobrov, K.~Smirnov, and G.~Chernishev, ``Desbordante: a
  framework for exploring limits of dependency discovery algorithms,'' in
  \emph{2021 29th Conference of Open Innovations Association (FRUCT)}, 2021,
  pp. 344--354.

\bibitem{10143047}
A.~Smirnov, A.~Chizhov, I.~Shchuckin, N.~Bobrov, and G.~Chernishev, ``Fast
  discovery of inclusion dependencies with desbordante,'' in \emph{2023 33rd
  Conference of Open Innovations Association (FRUCT)}, 2023, pp. 264--275.

\bibitem{szlichta2012fundamentals}
\BIBentryALTinterwordspacing
J.~Szlichta, P.~Godfrey, and J.~Gryz, ``Fundamentals of order dependencies,''
  \emph{Proc. VLDB Endow.}, vol.~5, no.~11, p. 1220–1231, jul 2012. [Online].
  Available: \url{https://doi.org/10.14778/2350229.2350241}
\BIBentrySTDinterwordspacing

\bibitem{DBLP:journals/ATDS/Wilfred01}
\BIBentryALTinterwordspacing
W.~Ng, ``An extension of the relational data model to incorporate ordered
  domains,'' \emph{ACM Trans. Database Syst.}, vol.~26, no.~3, p. 344–383,
  sep 2001. [Online]. Available: \url{https://doi.org/10.1145/502030.502033}
\BIBentrySTDinterwordspacing

\bibitem{DBLP:journals/VLDB/SzlichtaGGKS18}
\BIBentryALTinterwordspacing
J.~Szlichta, P.~Godfrey, L.~Golab, M.~Kargar, and D.~Srivastava, ``Effective
  and complete discovery of bidirectional order dependencies via set-based
  axioms,'' \emph{The VLDB Journal}, vol.~27, no.~4, p. 573–591, aug 2018.
  [Online]. Available: \url{https://doi.org/10.1007/s00778-018-0510-0}
\BIBentrySTDinterwordspacing

\bibitem{DBLP:journals/VLDB/ZijingASS20}
\BIBentryALTinterwordspacing
Z.~Tan, A.~Ran, S.~Ma, and S.~Qin, ``Fast incremental discovery of pointwise
  order dependencies,'' \emph{Proc. VLDB Endow.}, vol.~13, no.~10, p.
  1669–1681, jun 2020. [Online]. Available:
  \url{https://doi.org/10.14778/3401960.3401965}
\BIBentrySTDinterwordspacing

\bibitem{DBLP:journals/VLDB/XuIP13}
X.~Chu, I.~F. Ilyas, and P.~Papotti, ``Discovering denial constraints,''
  \emph{Proc. VLDB Endow.}, vol.~6, no.~13, p. 1498–1509, aug 2013.

\bibitem{DBLP:journals/VLDB/YifengLZ20}
Y.~Jin, L.~Zhu, and Z.~Tan, ``Efficient bidirectional order dependency
  discovery,'' in \emph{2020 IEEE 36th International Conference on Data
  Engineering (ICDE)}, 2020, pp. 61--72.

\bibitem{DBLP:journals/VLDB/HemantLI19}
H.~Saxena, L.~Golab, and I.~F. Ilyas, ``Distributed implementations of
  dependency discovery algorithms,'' \emph{Proc. VLDB Endow.}, vol.~12, no.~11,
  p. 1624–1636, jul 2019.

\bibitem{DBLP:journals/VPDB/SchmidlP21}
S.~Schmidl and T.~Papenbrock, ``Efficient distributed discovery of
  bidirectional order dependencies,'' \emph{The VLDB Journal}, vol.~31, no.~1,
  p. 49–74, aug 2021.

\bibitem{DBLP:journals/DMKD/YaoH08}
H.~Yao and H.~J. Hamilton, ``Mining functional dependencies from data,''
  \emph{Data Min. Knowl. Discov.}, vol.~16, no.~2, p. 197–219, apr 2008.

\bibitem{DBLP:journals/VLDB/WyssGR01}
C.~Wyss, C.~Giannella, and E.~Robertson, ``Fastfds: A heuristic-driven,
  depth-first algorithm for mining functional dependencies from relation
  instances extended abstract,'' in \emph{DaWaK}, Y.~Kambayashi \emph{et~al.},
  Eds.\hskip 1em plus 0.5em minus 0.4em\relax Berlin, Heidelberg: Springer
  Berlin Heidelberg, 2001, pp. 101--110.

\bibitem{DBLP:journals/TCJ/HuhtalaKPT99}
Y.~Huhtala, J.~Kärkkäinen, P.~Porkka, and H.~Toivonen, ``Tane: An efficient
  algorithm for discovering functional and approximate dependencies,''
  \emph{The Computer Journal}, vol.~42, no.~2, pp. 100--111, 1999.

\bibitem{DBLP:journal/ICDE/KaregarMGGKSS22}
R.~Karegar, M.~Mirsafian, P.~Godfrey, L.~Golab, M.~Kargar, D.~Srivastava, and
  J.~Szlichta, ``Discovering domain orders via order dependencies,'' in
  \emph{ICDE'22}, 2022, pp. 1098--1110.

\bibitem{DBLP:journals/edbt/ConsonniSMV19}
C.~Consonni \emph{et~al.}, ``Discovering order dependencies through order
  compatibility,'' in \emph{{EDBT'19}}, M.~Herschel \emph{et~al.}, Eds.\hskip
  1em plus 0.5em minus 0.4em\relax OpenProceedings.org, 2019, pp. 409--420.

\bibitem{godfrey2019errata}
P.~Godfrey, L.~Golab, M.~Kargar, D.~Srivastava, and J.~Szlichta, ``Errata note:
  Discovering order dependencies through order compatibility,'' \emph{CoRR},
  vol. abs/1905.02010, 2019.

\end{thebibliography}

\end{document}